**Bulk metallic glass-like structure of small icosahedral metallic nanoparticles**

Vicky V. T. Doan-Nguyen[1,*], Simon A. J. Kimber[2,*], Diego Pontoni[2], Benjamin T. Diroll[3], Danielle C. Reifsnyder[3], Marcel Miglierini[4,5], Xiaohao Yang[6], Christopher B. Murray[1,3], and Simon J. L. Billinge[6,7, †]

[1] *Department of Materials Science and Engineering, University of Pennsylvania, Philadelphia, Pennsylvania, 19104 USA;*

[2] *European Synchrotron Radiation Facility (ESRF), Grenoble Cedex 9, 38043, France;*

[3] *Department of Chemistry, University of Pennsylvania, Philadelphia, Pennsylvania, 19104 USA;*

[4] *Slovak University of Technology, Faculty of Electrical Engineering and Information Technology, Institute of Nuclear and Physical Engineering, Ilkovičova 3, 81219 Bratislava, Slovakia;*

[5] *RCATM, Palacky University, 17. listopadu 12, 771 46 Olomouc, Czech Republic;*

[6] *Department of Applied Physics and Applied Mathematics, Columbia University, New York, 10027 USA;*

[7] *Condensed Matter Physics and Materials Science Department, Brookhaven National Laboratory, Upton, New York, 11973 USA.*

[*]These authors contributed equally to this work.

**We demonstrate a remarkable equivalence in structure measured by total X-ray scattering methods between very small metallic nanoparticles and bulk metallic glasses (BMGs), thus connecting two disparate fields, shedding new light on both. Our results show that for nanoparticle diameters <5 nm the structure of Ni nanoparticles changes from *fcc* to the characteristic BMG-like structure, despite them being formed from a single element, an effect we call nano-metallic glass (NMG) formation. However, high-resolution TEM images of the NMG clusters exhibit lattice fringes indicating a locally well-ordered, rather than glassy, structure. These seemingly contradictory results may be reconciled by finding a locally ordered structure that is highly isotropic and we show that local icosahedral packing within 5 atomic shells explains this. Since this structure is stabilized only in the vicinity of a surface which highlights the importance of the presence of free volume in BMGs for stabilizing similar local clusters.**

The ability to make amorphous metals in the bulk form as bulk metallic glasses (BMG) surprised the scientific world and led to novel materials with amazing mechanical strength, toughness and magnetic properties[1–5]. They are made by rapidly cooling from the melt[4,5] a mixture of different sized elements that are not glass-formers in their elemental forms. Despite many studies, our understanding of the BMG structure and formation is far from complete[6–9]. Here, we demonstrate using total X-ray scattering methods a remarkable structural equivalence between very small metallic nanoparticles and a broad range of BMGs, thus establishing an unexpected connection between two disparate fields of study, shedding new light on both. Our results show that below a threshold diameter of ~5 nm, the structure of Ni nanoparticles changes from *fcc* close-packing to the characteristic BMG structure, an effect we call nano-metallic glass (NMG) formation. However, in common with other studies, high resolution TEM images of our NMG clusters exhibit lattice fringes indicating a locally well ordered, rather than glassy, structure[10,11]. These seemingly contradictory results can be reconciled by assigning to the particles an atomic structure that is highly isotropic and only locally ordered. We show that local icosahedral packing of up to 5 atomic shells can explain our X-ray results. In the NMGs, this icosahedral structure is evident in the smallest particles where it is always in the vicinity of a surface, suggesting the importance of the presence of free volume in BMGs for stabilizing similar local clusters. The pair distribution functions of larger metallic nanoparticles also reveal a minority NMG component, which presumably comes from the near-surface region, suggesting that the catalytically important nanoparticle surfaces may be substantially icosahedral in nature even for larger *fcc* nanoparticles.

Elemental metals typically form straightforward space-filling structures such as face centered cubic (*fcc*) or hexagonal close packed (*hcp*). Other low energy structures exist which are non-space filling and form in small clusters[10,12] but become unstable in the bulk because they imply the formation of internal voids and free volume[13]. Non-close-packed metallic structures can be kinetically trapped by rapid cooling, for example in the case of bulk metallic glasses (BMG), which form from highly tuned mixes of particular elements[5]. While the resulting amorphous structures have remarkable properties, the principles that dictate when and how they form, and details of the resulting atomic arrangements, are not completely understood. For example, the BMG structure has variously been proposed to show packing (possibly fractal) of small clusters or to result from a frustrated competition between different structures[14] though the X-ray diffraction patterns indicate an amorphous structure consistent with their glassy dynamics[15].

Another challenge is to understand the structure of metallic nanoparticles, which are typically between 1 and 100 nm in size. Metal nanoparticles, in particular, have excellent catalytic activity with applications such as Suzuki coupling[16], $NO_x$ reduction in exhaust streams[17], oxygen-reduction reaction on fuel cell cathodes[18,19], preferential oxidation of CO in hydrogen[20], or methane combustion[21]. Distinction is often made in the literature between clusters and colloids[22], where clusters are ultra-small assemblies of metallic ions coordinated by organic ligands. They resemble organometallic complexes and have unique and well-defined structures, which are generally determined from single crystals of the clusters. On the other hand, colloidal particles tend to be larger and have properties that more resemble quantum confined bulk metals than organometallic molecules, and despite notable exceptions[23], they tend not to crystallize making structure determination a challenge. Some structural information may be inferred from HRTEM images, and great progress has been made in categorizing certain classes of small metallic clusters by packing type[10]. However, as the Jadzinsky *et al*. study showed[23], the actual atomic arrangements may be complex and non-trivial, not simply platonic figures. Understanding and controlling the structure of these materials is of considerable economic importance.

Quantitative nanoscale structure determination requires bulk probes that yield structural information on length scales below 100 nm. Recent developments of hard x-ray total scattering and atomic pair distribution function (PDF) analysis have proven ideal for examining the structure of such nanoparticles[24, 25,26]. PDF studies complement high-resolution transmission electron microscopy (HRTEM) experiments, which yield atomically resolved images of individual nanoparticles, by providing high precision measurements of bond-length distributions and atomic arrangements averaged over the whole sample. Since the PDF is a sample average, the most precise structural information about individual nanoparticles requires samples that have great structure, size and shape uniformity, otherwise sample polydispersity may limit the information available in the PDF. In cases where X-ray methods do not yield sufficient information to constrain a unique structural solution[24], combining PDF data with complementary measurements such as HRTEM should be pursued.

In the current study we use nanoparticle samples prepared with precise control over size and shape uniformity, taking advantage of recent developments in metallic nanoparticle synthesis control. We characterize the particles by acquiring both high quality PDF and HRTEM data. The uniformity of our samples is demonstrated by TEM images of Ni particles (Fig. 1b). TEM images of the Pd and alloy samples (not shown) exhibit similar uniformity. The PDF analysis was carried out on high energy synchrotron X-ray data from size selected Ni, Pd, and $Ni_xPd_{1-x}$ alloy nanoparticles as a function of size and composition. HRTEM images were also collected for the smallest nanoparticles.

The PDF is the Fourier transform of the properly corrected and normalized total scattering powder diffraction intensity, and may be understood as a histogram of the atomic distances in the material[27]. For example, the nearest neighbor distance in *fcc* nickel is 2.7 Å corresponding to the first peak in the PDF. The PDFs of the larger particles show sharp peaks across a wide range of *r* as shown in Fig. 1a. These could be modeled[28] using the *fcc* structure of bulk nickel modified by a spherical envelope function responsible for the fall-off in PDF peak intensity with increasing-*r* due to the finite particle size. However, this model fails to reproduce our data for the 5 nm particles. For the smallest particles, the PDF peaks are extremely broad, reflecting significant disorder. Even with extensive peak broadening, the model peaks from the best-fit *fcc* model (continuous lines Fig. 1a) are completely out-of-phase in the high-*r* region and clearly do not reproduce the measured PDF, which therefore is not simply describing a disordered: it is not possible to fit self-consistently with an *fcc* model both the low-*r* and high-*r* peaks in the PDF. Similar results were found in all small-sized (< 5nm) NMGs throughout the entire solid solution of $Ni_xPd_{1-x}$ nanoparticles.

There is a striking resemblance between the PDFs of the small nanoparticles and those of bulk metallic glasses[7]. This is illustrated in Fig. 2, which includes the PDF of a macroscopic ribbon of $Fe_{76}MoCu_1B_{15}$ representing typical BMG structures found in the literature[29]. In Fig. 2a, the BMG PDF is over plotted by the PDF of 5 nm Pd nanoparticles with only a scale factor applied. The overall similarity of the curves is immediately apparent with broad sinusoidal features in both cases, although the phase and wavelength of the oscillations differ. However, a simple re-scaling of the distance axis brings the two curves into correspondence. In Fig. 2b we show the two curves plotted on top of each other after rescaling by the average metallic radius $<r_{metallic}>$ of each sample. This simple scaling yields an agreement between the two data sets that is remarkable and superior to the best-fit *fcc* model. This is a clear indication that the nanoparticles, despite being made from elemental Ni, are mainly characterized by BMG-like disorder rather than close-packed *fcc* structure. Whilst "disorder" has been observed in some small metallic nanoparticles[10], this close, quantitative, relationship to the BMG structure has not been noted to date.

The similarity between the nanoparticle and the rescaled BMG PDFs extends beyond peak positions and widths to the damping envelope, which is a measure of the range of structural coherence of the local order. The BMGs obtained from a wide range of compounds[2,30] exhibit universal PDF features. Our results extend such universality to simple-metal nanoparticles. The BMG structure has never been observed in pure elements, yet it appears as the preponderant structure in our small mono- and bimetallic nanoparticles. To differentiate the BMG-like nanoparticles from conventional BMG's we refer to them as nano-metallic glasses (NMGs).

We now explore the ubiquity of the NMG structure for a range of small Ni, Pd and $Ni_{1-x}Pd_x$ alloy nanoparticles. As shown in Fig. 3a, the similarity of the intermediate range structure in all the samples is clearly apparent, with a broadened and non-*fcc*-like PDF above ~6 Å. This PDF can be well fit with a damped single-mode sine wave as shown in the Figure. This fact suggests that in this intermediate range the structure is isotropic, distinct from close-packed structures such as *fcc* and *hcp* that have different periodicities in the different crystallographic directions. In order to test whether the NMG samples show the same $<r_{metallic}>$ scaling that we reported above, in Fig. 3b we show all the curves plotted on top of each other. This simple scaling works remarkably well for all the Ni and alloy samples, showing that the NMG state is robust to a large variation in mean metallic radius. The agreement is still good, but less perfect, for the pure Pd nanoparticles.

The discovery of an equivalence in PDF between BMG and NMG systems allows us to extend an analysis of the first-sharp diffraction peak (FSDP) scaling with atomic volume, $V_a$, which was performed by Ma *et al.*[9] on a large number of BMG structures. They reported a power-law scaling between the position of the FSDP, $Q_{FSDP}$, and $V_a$, with an exponent that suggested a fractal network. The Fourier relationship between the diffraction pattern and the PDF transforms the FSDP into a damped single-mode sine-wave in the PDF, precisely the oscillatory signal that we observe in Fig. 3, where the wavelength of the sine-wave, $\lambda_D$, is given by $\lambda_D = 2\pi/Q_{FSDP}$. We therefore replot in Fig. 4a the data from Ma *et al.*[9] as $\lambda_D$ vs. $<r_{metallic}>$, superimposing the values for the NMGs from the current study. The NMG samples add valuable points that extend the narrow range of $<r_{metallic}>$ from the earlier study. All the data lie on a straight line suggesting that $\lambda_D$ scales simply with $<r_{metallic}>$ (which is proportional to the cube root of $V_a$ used in the earlier study) without invoking fractal behavior. It is in fact difficult to justify a fractal structure in such small and chemically simple systems as the elemental NMGs. However, our results suggest that the intermediate range structure of both NMG and BMG materials presents the common feature of being highly isotropic as described by a single period density wave in all directions

Observing BMG-like behavior in nanoparticles allows us to investigate its origin further. Ultra-small metallic clusters often form in cuboctahedral, decahedral or icosahedral morphologies, as suggested by HRTEM measurements[10]. Although our clusters are somewhat larger, it is interesting to look for evidence in HRTEM images of similar behavior in the NMG clusters. HRTEM images of the 5 nm clusters are shown in Fig. 5, where it is immediately clear that, despite the "glassy" PDF signal, lattice fringes are evident in the images, suggesting a well-defined short-range structure. By tilting the microscope stage appropriately it is possible to see lattice fringes in virtually all the particles suggesting that this is not a minor effect affecting only a few particles. To reconcile these apparently contradictory HRTEM and PDF results we calculated the PDFs of icosahedral clusters. The calculated PDFs for 309-atom icosahedra, which consist of a core atom surrounded by 4 additional atomic shells, are shown in Fig. 5 with cartoon images of the icosahedral structure in the upper inset. The green curves are all PDFs calculated from this icosahedral model with different atomic displacement parameters (ADPs), which are a measure of the static and dynamic atomic disorder. The top curve has an unrealistically small ADP of 0.001 Å$^2$ but it serves to illustrate the large number of distinct atomic distances present in the perfect icosahedral cluster. The second green curve was calculated with an ADP of 0.01 Å$^2$, which is a reasonable value if there was just thermal motion but no static disorder in the material. The third green curve—overlaid on top of the 5 nm Ni PDF—was calculated with a large value of 0.1 Å$^2$. This is appropriate if there is a broad atomic positional distribution around the average sites in the icosahedral cluster. In the latter two cases a damped sine wave is shown overlaid in black. The good agreement shows that the PDF of a broadened icosahedral model is well represented by a damped single-mode sine wave, consistent with the measured PDF of the NMG clusters. Moreover, we have attempted fits of the icosahedral model to the NMG PDFs where the only tunable structural parameters are an isotropic breathing parameter that allows the cluster to uniformly shrink or expand, and a single global ADP. Additionally, the fits include a tunable scale factor and a correlation parameter that allows lower-$r$ PDF peaks to be sharper, thus bringing to 5 the total number of fitting parameters. The results of the fit are shown as the bottom curve in Fig. 5. The agreement between the fit and experimental data shows that a well-defined local icosahedral atomic geometry, albeit with a large atomic density distribution around each average atomic position, is consistent with the PDF data. The fringes in the TEM survive presumably because the intermediate range order is well defined. Even though the positional order of the atoms around each average atomic position is loose, the average position itself is well enough defined to yield interference and fringes in the HRTEM image.

The appearance of the NMG structure below a critical diameter in Ni suggests that the vicinity of a surface may be important to stabilize it. Atomic arrangements routinely reconstruct at surfaces[31] and highly curved surfaces may result in reconstructions that appear quite disordered[32]. However, in these nanoparticles the structure of the core and the surface modifies below a critical diameter, which is much larger than the first one or two atomic layers, but comparable to twice the range of coherence of the intermediate order. This means that the icosahedral cluster around any origin atom will not span the diameter of the particle, but is likely to impinge on a surface in some direction. We speculate that the large ADP values may arise from the need to create interpenetrating icosahedra, centered on different atoms, that are not completely compatible with each other. In this picture the nanoparticles apparently prefer defective icosahedral packing to *fcc* order at these small sizes. We speculate that the alloying of elements with a large size distribution in BMGs achieves a similar effect, as packing mismatches create voids and free volume in the system, allowing the non-space filling, possibly icosahedral, packing to form, with small interstitial atoms filling the voids to lower the energy further. Such ideas are not new[14] but this work gives direct experimental support to their validity.

The idea that the presence of a surface might stabilize the NMG state led us to explore the possibility that there is a thin layer of NMG state at the surface of the larger Ni nanoparticles. Close examination of the difference curves in Fig. 1 indicates that after fitting the *fcc* model there is considerable signal left in the residual which, although noisy, appears to have an oscillatory nature. The difference curves are plotted in Fig. 4b and compared to the PDF of the 5 nm Ni particles. The similarity is again striking, and the same Gaussian damped single-mode sine wave fits the data well. These observations are highly suggestive of a NMG component in the larger particles, which we speculate is at the surface.

In summary, we have shown that an NMG structure, highly analogous to the structure of bulk metallic glasses, emerges in very small Ni and Pd nanoparticles. We show that this structure is consistent with the presence of an average geometry of 4 or 5 shell icosahedral clusters with significant atomic smearing. The stabilization of this non-space-filling cluster in very small nanoparticles, in the presence of a nearby surface, suggests the importance of the presence of free volume in stabilizing the structure in the bulk. The likelihood of an NMG layer on the surface of larger Ni particles has profound implications for catalysis employing supported nanoparticles, since heterogeneous catalysis is mediated by the surface.

**Supplementary Information** is linked to the online version of the paper at http://www.nature.com/naturecommunications.

**Acknowledgements** We thank the University of Pennsylvania Nano-Bio Interface Center and MINATEC Exchange program under the U.S. National Science Foundation, Office of International Science and Engineering, Division of International Research Experience for Students Award Number 1130994 for funding the visit of VDN to Grenoble. VDN and CBM are also supported by ARPA-E Award Number DE-AR0000123 for development of Ni and Pd samples. BTD, DCR, and CBM acknowledge support from the U.S. Department of Energy, Office of Basic Energy Sciences, Division of Materials Science and Engineering, under Award Number DE-SC0002158 for SAXS and HRTEM data collection. DCR acknowledges additional support from the NSF-IGERT program (Grant DGE-0221664). This work was performed in part at NCEM, which is supported by the Office of Science, Office of Basic Energy Sciences of the U.S. Department of Energy under Contract No. DE-AC02-05CH11231. We thank Chengyu Song for assistance at NCEM. The European Synchrotron Radiation Facility is acknowledged for access to instrumentation. The PDF data analysis and modeling in the Billinge Group was supported as part of a Brookhaven National Laboratory (BNL) Laboratory Directed Research and Development grant. BNL is supported by U.S. Department of Energy, Office of Science, Office of Basic Energy Sciences under Award Number DE-AC02-98CH10886. MM acknowledges the grant No. CZ.1.07/2.3.00/20.0155.

**Author Contributions** Nanoparticle samples were prepared and characterized by TEM and ICP-OES by VDN. SAXS was done and analyzed by VDN, SAJK, DP, and BTD. The BMG sample was provided by MM. Pair distribution function measurements were performed by SAJK and VDN. DCR performed HRTEM at NCEM. XY carried out PDF modeling. CBM provided input for the manuscript and general advice. The results were interpreted by VDN, SAJK, and SJLB, who wrote the paper.

**Author Information** Reprints and permissions information is available at http://www.nature.com/reprints. We declare no competing financial interests. Correspondence and requests for materials should be addressed to SAJK (kimber@esrf.fr) and SJLB (sb2896@columbia.edu).

**Methods Summary**

*Synthesis:* Nickel(II) acetylacetonate, palladium(II) acetylacetonate, oleylamine (70%), trioctylphosphine (97%), and 1-octadecene were purchased from Sigma Aldrich. Synthesis of nanoparticles involves Schlenk line techniques that utilized thermal decomposition of a metal-salt precursor in a flask with surfactants and a high-boiling solvent. The reactant amounts and reaction and purification conditions are detailed in SI Table 1.

*Characterization*: The nanoparticles were deposited on 300-mesh carbon-coated copper grids purchased from Electron Microscopy Sciences. TEM was done on a JEOL 1400 TEM with a $LaB_6$ filament, operating at 120 kV and equipped with an SC1000 ORIUS CCD camera and Digital Micrograph software. HRTEM was done on a JEOL 2010F TEM/STEM, equipped with a field emission gun (FEG), operating at 200 kV as well as the National Center for Electron Microscopy's Philips CM300FEG/UT TEM with an FEG and low spherical aberration ($C_s$ = 0.60 mm), operating at 300 kV. For ICP-OES, the nanoparticle samples were digested in 69.4% $HNO_3$ for 24 hours. The solutions were diluted to 0.7-7 ppm. The nickel and palladium calibration standards (0.1-500 ppm) were prepared by diluting from Sigma Aldrich TraceCERT-grade stock solutions of nickel 1000 ppm and palladium 970 ppm. The measurements were done using a Spectro Genesis spectrometer. Small-angle X-ray scattering data was collected at Penn using a Multi-Angle X-ray Scattering system with 1.54 Å X-ray wavelength with detector distances at 11 and 54 cm as well as at the ESRF at ID02 using 12.5 keV X-rays with detector distances at 1.5 and 10 m covering a *q* range of 0.5-0.001 Å$^{-1}$. Collection of the X-ray scattering data was done at the ESRF at ID15B using 87.8 keV X-rays (0.1412 Å) and a Mar345 detector. The raw images were integrated using Fit2D[33]. Background contributions from the 2 mm Kapton capillary tubes as well as Compton and fluorescence contributions were subtracted from the data. An in-house code 'iPDF' written by SAJK was used to correct and Fourier transform the data into real space pair distribution functions. PDFgui was used to model the PDFs in the small-box approximation[28].

**References**

1. Johnson, W. Fundamental aspects of bulk metallic glass formation in multicomponent alloys. *Materials Science Forum* **225-227**, 35–50 (1996).

2. Wang, W. H., Dong, C. & Shek, C. H. Bulk metallic glasses. *Materials Science and Engineering: R: Reports* **44**, 45–89 (2004).

3. Inoue, A. & Takeuchi, A. Recent development and application products of bulk glassy alloys✩. *Acta Materialia* **59**, 2243–2267 (2011).

4. Kumar, G., Tang, H. X. & Schroers, J. Nanomoulding with amorphous metals. *Nature* **457**, 868–72 (2009).

5. Peker, A. & Johnson, W. L. A highly processable metallic glass: Zr41.2Ti13.8Cu12.5Ni10.0Be22.5. *Applied Physics Letters* **63**, 2342 (1993).

6. Miracle, D. B. A structural model for metallic glasses. *Nature Materials* **3**, 697–702 (2004).

7. Hwang, J. *et al.* Nanoscale Structure and Structural Relaxation in Zr50Cu45Al5 Bulk Metallic Glass. *Physical Review Letters* **108**, 195505 (2012).

8. Zeng, Q. *et al.* Long-range topological order in metallic glass. *Science (New York, N.Y.)* **332**, 1404–6 (2011).

9. Ma, D., Stoica, A. D. & Wang, X. L. Power-law scaling and fractal nature of medium-range order in metallic glasses. *Nature Materials* **8**, 30–4 (2009).

10. Marks, L. D. Experimental studies of small particle structures. *Reports on Progress in Physics* **603**, 603–649 (1994).

11. Rodriguez-Lopez, J. L., Montejano-Carrizale, J. M. & Jose-Yacaman, M. Low dimensional non-crystallographic metallic nanostructures: HRTEM simulation, models and experimental results. *Modern Physics Letters B* **20**, 725–751 (2006).

12. Frenkel, A. Solving the 3D structure of metal nanoparticles. *Zeitschrift für Kristallographie* **222**, 605–611 (2007).

13. Voloshin, V. P. & Naberukhin, Y. I. On the origin of the splitting of the second maximum in the radial distribution function of amorphous solids. *Journal of Structural Chemistry* **38**, 62–70 (1997).

14. Cheng, Y. Q. & Ma, E. Atomic-level structure and structure–property relationship in metallic glasses. *Progress in Materials Science* **56**, 379–473 (2011).

15. Busch, R., Kim, Y. J. & Johnson, W. L. Thermodynamics and kinetics of the undercooled liquid and the glass transition of the Zr41.2Ti13.8Cu12.5Ni10.0Be22.5 alloy. *Journal of Applied Physics* **77**, 4039 (1995).

16. Kim, S.-W., Kim, M., Lee, W. Y. & Hyeon, T. Fabrication of hollow palladium spheres and their successful application to the recyclable heterogeneous catalyst for suzuki coupling reactions. *Journal of the American Chemical Society* **124**, 7642–3 (2002).

17. Zhou, S., Varughese, B., Eichhorn, B., Jackson, G. & McIlwrath, K. Pt-Cu core-shell and alloy nanoparticles for heterogeneous NO(x) reduction: anomalous stability and reactivity of a core-shell nanostructure. *Angewandte Chemie (International ed. in English)* **44**, 4539–43 (2005).

18. Stamenkovic, V. R. *et al.* Trends in electrocatalysis on extended and nanoscale Pt-bimetallic alloy surfaces. *Nature Materials* **6**, 241–247 (2007).

19. Wu, J. *et al.* Icosahedral platinum alloy nanocrystals with enhanced electrocatalytic activities. *Journal of the American Chemical Society* **134**, 11880–3 (2012).

20. Alayoglu, S., Nilekar, A. U., Mavrikakis, M. & Eichhorn, B. Ru-Pt core-shell nanoparticles for preferential oxidation of carbon monoxide in hydrogen. *Nature Materials* **7**, 333–338 (2008).

21. Cargnello, M. *et al.* Exceptional activity for methane combustion over modular Pd@CeO2 subunits on functionalized Al2O3. *Science* **713**, 713–717 (2013).

22. Lewis, L. Chemical catalysis by colloids and clusters. *Chemical Reviews* **93**, 2693–2730 (1993).

23. Jadzinsky, P. D., Calero, G., Ackerson, C. J., Bushnell, D. a & Kornberg, R. D. Structure of a thiol monolayer-protected gold nanoparticle at 1.1 A resolution. *Science (New York, N.Y.)* **318**, 430–3 (2007).

24. Billinge, S. J. L. & Levin, I. The problem with determining atomic structure at the nanoscale. *Science (New York, N.Y.)* **316**, 561–5 (2007).

25. Neder, R. B. *et al.* Structural characterization of II-VI semiconductor nanoparticles. *Physica Status Solidi (C)* **4**, 3221–3233 (2007).

26. Tyrsted, C. *et al.* Understanding the formation and evolution of ceria nanoparticles under hydrothermal conditions. *Angewandte Chemie (International ed. in English)* **51**, 9030–3 (2012).

27. Egami, T. & Billinge, S. J. L. *Underneath the Bragg Peaks: Structural Analysis of Complex Materials*. (Elsevier: Amsterdam, 2013).

28. Farrow, C. L. *et al.* PDFfit2 and PDFgui: computer programs for studying nanostructure in crystals. *Journal of Physics: Condensed Matter* **19**, 335219 (2007).

29. Pan, S. P., Qin, J. Y., Wang, W. M. & Gu, T. K. Origin of splitting of the second peak in the pair-distribution function for metallic glasses. *Physical Review B* **84**, 1–4 (2011).

30. Schroers, J. Bulk metallic glasses. *Physics Today* **66**, 32 (2013).

31. Robinson, K. & Tweet, D. J. Surface x-ray diffraction. *Reports on Progress in Physics* **55**, 599–651 (1992).

32. Nam, H.-S., Hwang, N., Yu, B. & Yoon, J.-K. Formation of an icosahedral structure during the freezing of gold nanoclusters: Surface-induced mechanism. *Physical Review Letters* **89**, 275502 (2002).

33. Hammersley, A. P., Svensson, S. O., Hanfland, M., Fitch, A. N. & Hausermann, D. Two-dimensional detector software: From real detector to idealised image or two-theta scan. *High Pressure Research* **14**, 2350248 (1996).

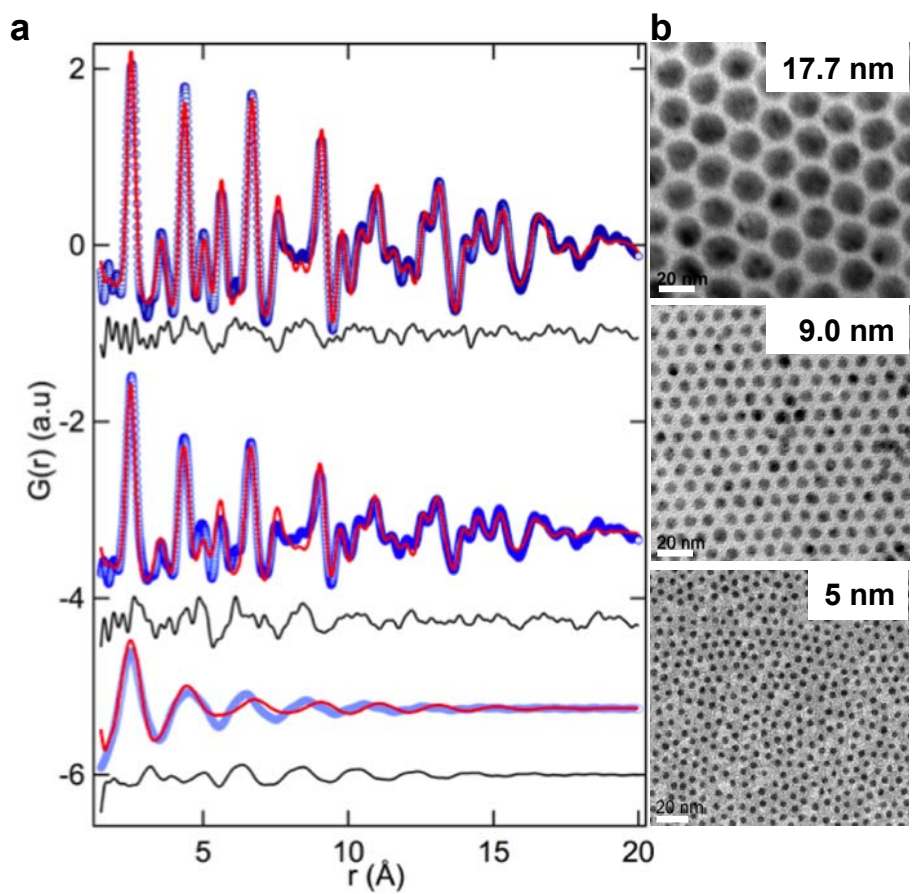

**Figure 1: Structural characterization of size controlled nickel nanoparticles. a**, Pair distribution functions for the three samples, as determined by Fourier transformation of high energy X-ray scattering data. The blue points represent the data, and the red lines are fits of *fcc* type models. The residuals of the fits are shown as black lines. **b**, Transmission electron microscopy images of nickel nanoparticles, note the formation of well-defined superlattices, which indicates uniformity. The scale bars correspond to 20 nm.

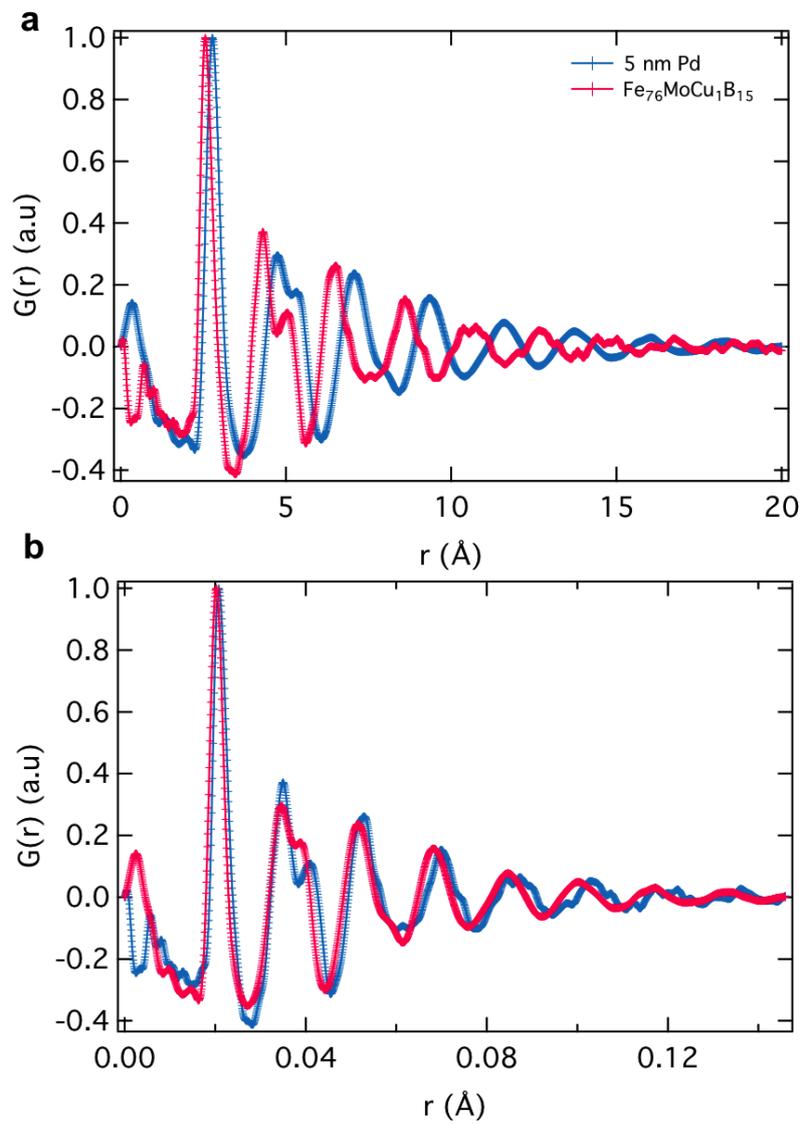

**Figure 2: Comparison of pair distribution functions of Pd nanoparticles and a representative bulk metallic glass, $Fe_{76}MoCu_1B_{15}$.** **a**, Comparison of the pair distribution functions after correction for an overall scale factor. **b**, Comparison of the pair distribution functions after scaling for the different metallic radii present in each sample.

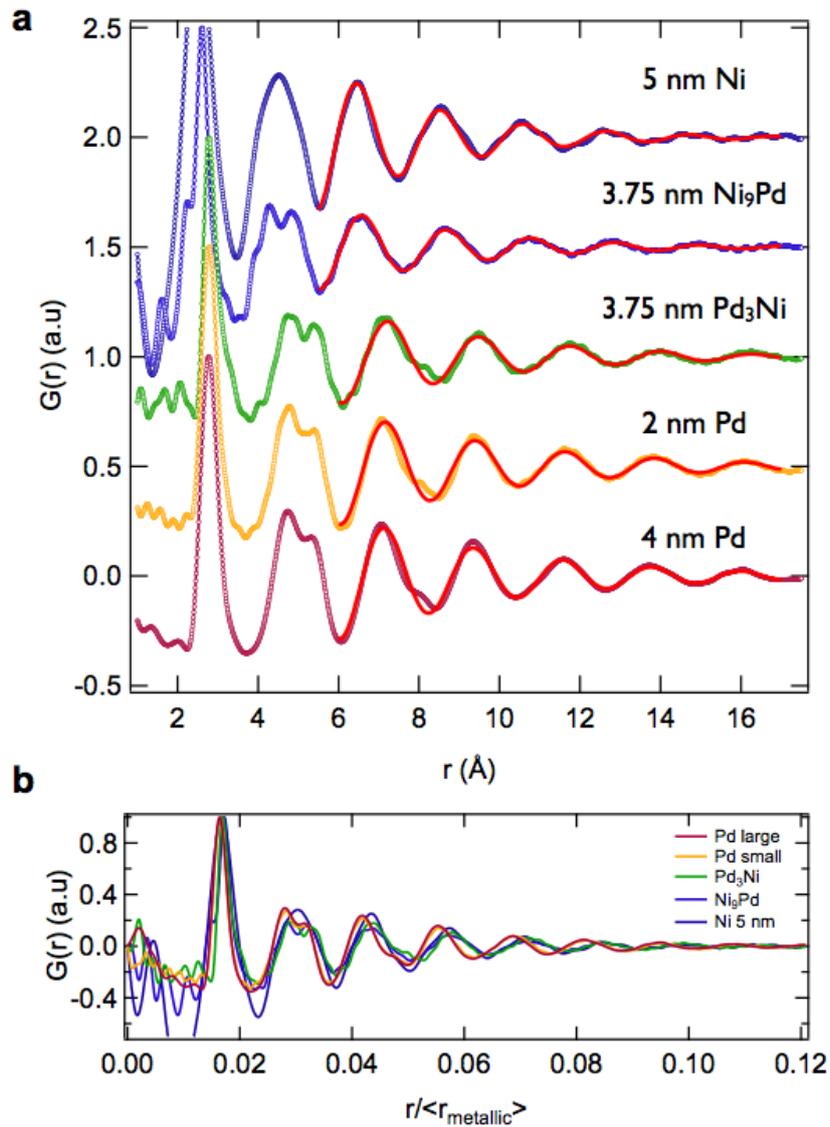

**Figure 3: Ubiquity of nano-metallic glass formation in the $Ni_{1-x}Pd_x$ solid solution series a**, Pair distribution functions of the solid solutions. The red lines show fits to the high-*r* region of the PDF of a Gaussian damped single-mode Sine function as described in the text. **b**, Comparison of the pair distribution functions after scaling by the different metallic radii in each sample.

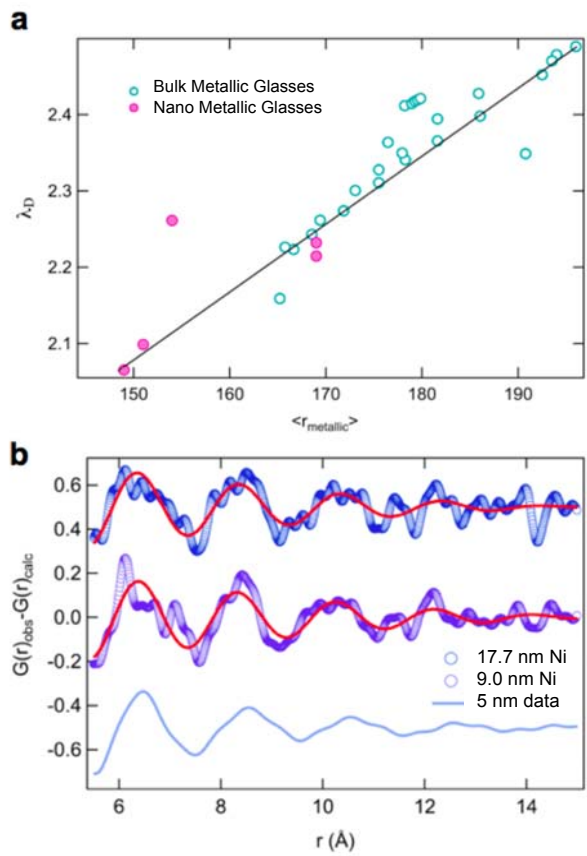

**Figure 4: Scaling of nano and bulk metallic glasses and evidence for a surface contribution in larger Ni particles a**, Linear dependence of the density wave fluctuations in NMGs and BMGs on the average metallic radii. Data points were extracted using the fits shown in Fig. 3a and from Ma *et al.*[9] **b**, Residuals from the fits of an *fcc* model to the data for larger Ni particles compared to the data for the 5 nm particles. Red lines show fits using a damped Sine wave function as above.

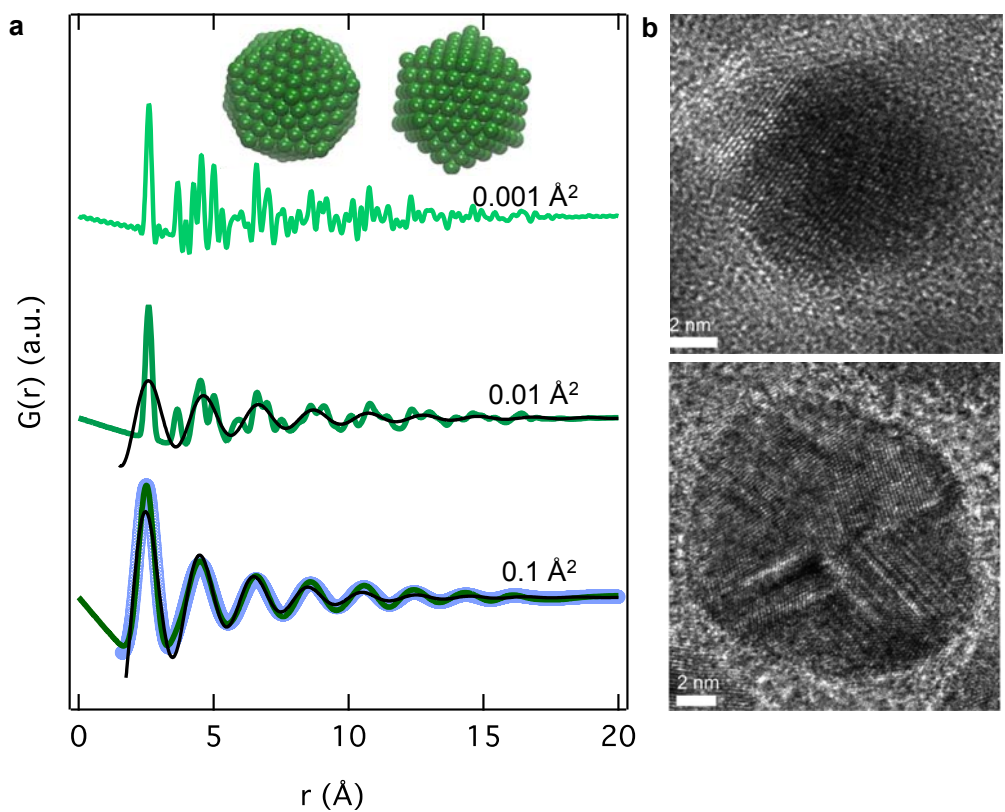

**Figure 5: PDFs of icosahedral models**, **a**. Green curves are all PDFs from the 309-atom icosahedral model shown in the inset. In the top curve the PDF was calculated with an unrealistically small ADP of 0.001 Å$^2$. A more reasonable value for the ADP in the absence of static disorder is 0.01 Å$^2$, which is shown in the second curve from the top. The bottom green curve is the same model calculated with an ADP of 0.1 Å$^2$, which implies a considerable non-thermal distribution of atomic positions around the average site. The black curves are fits of a damped single-mode sine wave to the icosahedral PDFs. The underlying blue curve at the bottom is the measured PDF from the 5 nm Ni nanoparticles. In this case, the single-mode sine wave is the best-fit PDF of the same icosahedral model where the only tunable parameters for refinement were a stretching parameter that allows the cluster to increase and decrease uniformly in diameter, a scale factor and a single ADP parameter applied to all the atoms, plus a "delta2" parameter[28] that sharpens the PDF peaks in the low-*r* region. **b**. High-resolution TEM images of 5 nm (top) and 17.7 nm (bottom) Ni nanoparticles show lattice fringes indicating existence of twinning and local ordering within the nanostructures.